# BOLTZMANN ENTROPY : PROBABILITY AND INFORMATION

C. G. Chakrabarti[1] and I. Chakrabarty[2]

[1]Department of Applied Mathematics, Calcutta University, Kolkata – 700 009, India
E-mail : cgc-math@rediflmail.com.
[2]Department of Mathematics, Heritage Institute of Technology, Kolkata – 700 107, India
E-mail : indranilc@indiainfo.com.

We have presented first an axiomatic derivation of Boltzmann entropy on the basis of two axioms consistent with two basic properties of thermodynamic entropy. We have then studied the relationship between Boltzmann entropy and information along with its physical significance.

**PACS** : 89.70, + C

## 1. INTRODUCTION

The concept of entropy was first introduced in thermodynamics by Clausius through the second law of thermodynamics. It was Boltzmann who first provided the statistical analogue of thermodynamic entropy linking the concept of entropy with molecular disorder or chaos [1]. The concept of entropy was later developed with the advent of statistical mechanics and information theory [3]. Boltzmann entropy is a building block to the foundation of statistical mechanics and the literature on Boltzmann entropy is vast. Inspite of this, there is scope for further extension of the existing methods and principles.

In the present paper we have provided first an axiomatic derivation of Boltzmann entropy on the basis of two axioms consistent with the additivity and the increasing law of thermodynamic entropy. We have then introduced the probability distribution of a macrostate through the concept of prior probabilities. This has led to an interesting relation connecting negentropy and information. The physical significance of this relation in the characterization of statistical equilibrium has been investigated.

## 2. THERMODYNAMIC PROBABILITY AND BOLTZMANN ENTROPY

Boltzmann entropy is defined by [1]
$$S = k \ln W \qquad \ldots \qquad (2.1)$$
where k is the thermodynamic unit of the measurement of the entropy and is the Boltzmann constant, W called the thermodynamic probability or statistical weight is the total number of microscopic states or complexions compatible with the macroscopic state of the system. Following Carnop [4], we shall, however, call W as the degree of disorder or simply disorder of the system. We can thus restate Boltzmann principle as
$$\text{Entropy} = k \ln (\text{Disorder}) \qquad \ldots \qquad (2.2)$$
The entropy S being a monotonic increasing function of disorder, can itself be considered as a measure of disorder of the system. Boltzmann entropy thus provides a measure of the disorder of the distribution of the states over permissible microstates [4]. Let us derive the expression (2.1) of entropy axiomatically on the basis of two fundamental properties of thermodynamic entropy. We make the following axioms :

**Axiom (i)** : The entropy S(W) of system is a positive increasing function of the disorder W, that is,

$$S(W) \leq S(W+1) \qquad \ldots \qquad (2.3)$$

This is motivated by the increasing law of entropy in thermodynamics.

**Axiom (ii)** : The entropy S(W) is assumed to be an additive function of the disorder W, that is, for any two statistically independent systems with degrees of disorder $W_1$ and $W_2$ respectively, the entropy of the composite system is given by

$$S(W_1 \cdot W_2) = S(W_1) + S(W_2) \qquad \ldots \qquad (2.4)$$

This axiom is the statistical analogue of the thermodynamic postulate of additivity of entropy. We have then the theorem :

**Theorem (1)** : If the entropy function S(W) satisfies the above two axioms (i) and (ii), then S(W) is given by

$$S = k \ln W \qquad \ldots \qquad (2.5)$$

where k is a positive constant depending on the unit of measurement of entropy.

**Proof** : Let us assume that $W > e$. This is justified by the fact that the macroscopic system we are interested in consists of a large number of microscopic states and hence corresponds to a large value of statistical weight or thermodynamic probability W. For any integer n, we can find an integer m(n) such that

$$e^{m(n)} \leq W^n \leq e^{m(n)+1} \qquad \ldots \qquad (2.6)$$

$$\text{or} \quad \frac{m(n)}{n} \leq \ln W \leq \frac{m(n)+1}{n}$$

$$\text{consequently,} \quad \lim_{n \to \infty} \frac{m(n)}{n} = \ln W \qquad \ldots \qquad (2.7)$$

The entropy function S(W) satisfies both the axioms (i) and (ii). By axiom (i) we have then from (2.6)

$$S(e^{m(n)}) \leq S(W^n) \leq S(e^{m(n)+1}) \qquad \ldots \qquad (2.8)$$

Again by axiom (ii) we have from (2.8)

$$m(n) S(e) \leq n S(W) \leq (m(n)+1) S(e)$$

$$\text{consequently,} \quad \lim_{n \to \infty} \frac{m(n)}{n} = \frac{S(W)}{S(e)} \qquad \ldots \qquad (2.9)$$

Comparing (2.7) and (2.9), we get

$$S(W) = k \ln W \qquad \ldots \qquad (2.10)$$

where $S(e) = k$ is a positive constant, the positivity follows from the positivity of the entropy – function postulated in axiom (i). The above theorem provides a rigorous derivation of the entropy function which is independent of the micro-model of the system – classical or quantal. In most of the books on statistical physics except a few [6] the expression of entropy (2.1) is determined from the additivity postulate of entropy only. This is not correct. The above derivation based on the axioms (i) and (ii) which are variant forms of two basic properties of thermodynamic entropy is more sound and is valid for macroscopic system containing a large number of molecules [6].

### 3. PRIOR PROBABILITIES: ENTROPY AND INFORMATION

The thermodynamic probability which appears in Boltzmann entropy (2.1) is not a probability but an integer. How to introduce the concept of probability ? It was Einstein [7] who first introduced the concept of probability of a macrostate by inverting Boltzmann principle [8]. We shall, however, follow a different approach. We consider an isolated system consisting of N molecules classified into n energy-states $\varepsilon_i$ ( i = 1, 2, ….n) and let $N_i$ ( i = 1, 2, ….n) be the occupation number of the ith energy-state $\varepsilon_i$. The macroscopic state of the system is given by the set of occupation numbers $A_n = \{N_1, N_2, … N_n\}$.

We are now interested in the probability distribution of the macrostate $A_n$ or the probability distribution of the set of occupation numbers $A_n = \{N_1, N_2, …, N_n\}$. In view of the many-body aspect of the system, the set of occupation numbers $\{N_1, N_2, … N_n\}$ can be assumed to be a set of random variables. The probability distribution of the macrostate $A_n = \{N_1, N_2, … N_n\}$ is then given by

$$P(A_n) = P(N_1, N_2, … N_n) = W(A_n) \prod_{i=1}^{n} p_i^{0^{N_i}} \quad … \quad (3.1)$$

where $p_i^0$ is the prior probability that a molecule lies in the ith energy-state $\varepsilon_i$. The probability (3.1) is of fundamental importance in connecting the disorder $W(A_n)$ with the probability $P(A_n)$, while the former is related to the Boltzmann entropy and the later to information as shall see. The appearance of prior probabilities $\{p_i^0\}$ makes the system complex. If no state is preferable to other, it is then reasonable to assume that all prior probabilities are equal to one another so that $p_i^o = 1/n$, (i = 1, 2, … n). This is Laplace principle of insufficient knowledge [10]. According to Jaynes [3] it is the state of maximum prior ignorance of the system. It that case it is easy to see from (3.1) that

$$- k \ln P(A_n) = S_{max} – S \quad … \quad (3.2)$$

where $S_{max} = k N \ln n$ is the maximum value of the entropy S. The l.h.s. of (3.2) is the information-content of the macrostate $A_n$. that is, the information gained from the realization of the macrostate $A_n$ [2]. The r.h.s of (3.2) is the negentropy of the system giving a measure of distance of non-equilibrium state from the equilibrium state [9, 12]. The relation (3.2) then implies the equivalence of information with negentropy consistent with Brillouin's negentropy principle of information [9].

Let us now investigate the significance of the relation (3.2) for the statistical equilibrium of a system. According to Boltzmann and Planck the statistical equilibrium of a system corresponds to the most probable state of the system. The thermodynamic probability $W(A_n)$ is not a probability, it is an integer. So the statistical equilibrium as assumed to correspond to the maximum of the thermodynamic probability or equivalently the Boltzmann entropy gives rise to some confusion [8]. However, from the relation (3.2) we see that the most probable state of the system obtained by maximizing the probability of the macrostate $P(A_n)$ is equivalent to the maximization of the Boltzmann entropy (2.1) or the maximization of the thermodynamic probability $W(A_n)$ provided we accept the hypothesis of equal a prior probability $p_i^o = 1/n$, (i = 1, 2, …n). It is in this way the confusion about the non-probabilistic character of thermodynamic probability $W(A_n)$ and the most-probable interpretation of statistical equilibrium can be resolved.

## 4. CONCLUSION

The object of the present paper is to present significantly different approach to the characterization of Boltzmann entropy in relation to both thermodynamic probability and information. The main results of the paper are as follows:

(i)   We have presented a rigorous axiomatic derivation of Boltzmann entropy on the basis of the axioms of additivity and increasing law of entropy consistent with two basic properties of thermodynamic entropy [11]. The method is superior, both mathematically and physically, to the existing methods (except a few) based mainly on the property of additivity of entropy only.

(ii)  Boltzmann entropy is independent of probability. We have introduced the probability of a macrostate through the hypothesis of prior probability $\{p_i^0\}$.

(iii) The probability distribution of macrostate leads to the interesting relation (3.2) connecting information and Boltzmann entropy or negentropy consistent with Brillouins' negentropy principle of information [4]. There is, however, a difference. In our case information comes through the concept of probability and the l.h.s of (3.2) is in fact, a probabilistic measure of information and not like the one introduced by Brillouin [9] and others [12].

(iv)  The relation (3.2) has important physical significance for statistical equilibrium. It resolves the confusion about the most–probable state of statistical equilibrium achieved by maximizing the thermodynamic probability under the hypothesis of equal prior probabilities.

(v)   In the case of unequal prior probabilities $\{p_i^o\}$ the situation becomes complicated and interesting for both physical and non-physical systems [13-15].